\documentclass[prl,aps,groupedaddress,showpacs,floatfix]{revtex4}

\usepackage{graphicx,amssymb,latexsym,amsmath,amsbsy,bm}

\newcommand{\etal}{{\it et al.,\;}}
\newcommand{\beq}{\begin{equation}}
\newcommand{\eeq}{\end{equation}}
\newcommand{\bea}{\begin{eqnarray}}
\newcommand{\eea}{\end{eqnarray}}

\newcommand{\veps}{\varepsilon}
\newcommand{\tr}{\mathrm{Tr}}

\newcommand{\benn}{\begin{displaymath}}
\newcommand{\eenn}{\end{displaymath}}
\newcommand{\EDC}{\textrm{EDC}}

\begin{document}

\title{Onset of a Pseudogap Regime in Ultracold Fermi Gases}
\author{ Piotr Magierski$^{1,2}$, Gabriel Wlaz{\l}owski$^{1}$ and Aurel Bulgac$^{2}$}

\affiliation{$^{1}$Faculty of Physics, Warsaw University of Technology,
ulica Koszykowa 75, 00-662 Warsaw, POLAND }
\affiliation{$^{2}$Department of Physics, University of Washington, Seattle,
WA 98195--1560, USA}

\begin{abstract}
We show, using an {\it ab initio} approach based on Quantum Monte Carlo technique, that the pseudogap regime emerges in ultracold Fermi gases close to the unitary point. We locate the onset of this regime at a value of the interaction strength corresponding to $(k_{F}a)^{-1}\approx -0.05$ ($a$ - scattering length). We determine the evolution of the gap as a function of temperature and interaction strength in the Fermi gas around the unitary limit and show that our results exhibit a remarkable agreement with the recent wave-vector resolved radio frequency spectroscopy data. Our results indicate that a finite temperature structure of the Fermi gas around unitarity is more complicated and involves the presence of the phase with preformed Cooper pairs, which however do not contribute to the long range order.
\end{abstract}

\date{\today}



\maketitle

The unitary Fermi gas is a dilute but an exceptionally strongly correlated ensemble of particles exhibiting universal properties and also a superfluid with a very large critical temperature, making it relevant to a wide range of systems including the quark-gluon plasma, neutron stars, nuclei \cite{schafer}, and to a certain extent to high Tc superconductors. A number of fundamental properties of the Fermi gas close to the unitary limit remain unknown and still pose an unprecedented challenge for the theory \cite{review1}. Although a convincing experimental proof of superfluidity was provided 6 years ago \cite{zwierlein2005}, the measurements of the pairing gap at finite temperatures are still in their infancy and only recently the technique  of photoemission spectroscopy allowed to probe the low-energy excitations of the gas at finite temperatures \cite{stewart2010}. The pairing gap measurements are believed to answer one of the most intriguing questions related to the strongly interacting Fermi gas: Does the simple picture of the phase transition from superfluid to normal state still holds around the unitary regime? Here we show, using an {\it ab initio} approach based on Quantum Monte Carlo (QMC) technique, that the pseudogap regime emerges in ultracold Fermi gases close to the unitary point. In the pseudogap regime fermions still have a strong tendency to be correlated with each other and behave to some extent as a system of bosons. Consequently above the critical temperature the system exists in an exotic state which is neither entirely normal nor superfluid and thus clearly eludes the understanding within the framework of classical BCS theory of superconductivity. 
This situation can be better understood from the viewpoint of the deep Bose-Einstein Condensation (BEC) 
limit, where the pairs of fermions form bound bosonic dimers. This implies 
the existence of two distinct temperature scales: the first one, $T_{c}$, related to the Bose-Einstein 
condensation of bosonic molecules and the second one, $T^{*}>T_{c}$, which is a temperature needed to break 
up a dimer into fermions. Somewhere around the unitary regime these two temperatures become 
practically indistinguishable and eventually merge into the critical temperature for superfluid to normal 
phase transition \cite{randeria}. The region between $T_{c}$ and $T^{*}$ is usually referred to as a pseudogap phase, where the system, 
although not being a superfluid, still exhibits the gap in the spectrum of quasiparticle excitations. 
Actually the term ``pseudogap phase'' may be somewhat misleading as, first, it is a strong depletion of the density of states 
in the spectrum and second, it is strictly speaking a particular regime, not a phase, as there is no phase transition between
pseudogap regime and the normal Fermi gas.

\begin{figure}[htb]
\includegraphics[width=6.5cm]{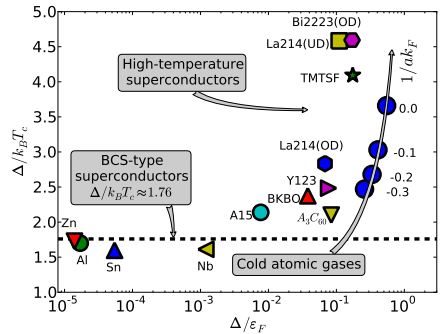}
\caption{ (Color online) Ratio of the pairing gap and the critical temperature for various superconductors (data extracted from Ref. \cite{fischer}). The clearly visible deviation from the typical ratio predicted by the BCS theory is a hallmark of HTSC. The same ratio for atomic Fermi gases close to the unitary limit are denoted by blue circles (data from Ref. \cite{bcs-bec}). \label{fig:A} }
\end{figure}

The situation in the ultracold atomic gases at the unitary regime is somewhat similar to that encountered in 
high temperature superconductors (HTSC) \cite{chen}, although the pseudogap regime in the cuprates
is not exclusively related to Cooper-pair formation \cite{kondo}. 
To appreciate these similarities one has to consider the ratio between 
the critical temperature and the pairing gap at zero temperature. According to the celebrated BCS theory 
this ratio is approximately equal to $1.76$ (for the temperature expressed in energy units, $k_B=1$). 
Any deviation from this value is an indication of the non-BCS superfluidity in the system. The origin of the deviation is usually related to the interaction strength and manifests itself in a rather large ratio of the pairing gap $\Delta$ to 
Fermi energy $\varepsilon_{F}$. It stands in contradiction to the implicit assumption of the BCS theory that 
the interaction is weak which imply $\Delta/\varepsilon_{F} \ll 1$. Clearly both ultracold gases and HTSC 
violate this assumption, see Fig.  \ref{fig:A}. Morevover the cold fermionic atoms constitute the system with the 
largest ratio $\Delta/\varepsilon_{F}$ and known HTSC only approach this limit \cite{luba}.

Theoretically the single particle gap in the fermionic spectrum can be determined from the spectral weight function $A(\bm{p},\omega)$, which carry information about the single particle excitation spectrum. 
In our approach we have used the Path Integral Monte Carlo (PIMC) technique on the lattice, which provides numerically essentially exact results with quantifiable uncertainties. The details of the approach have been described elsewhere \cite{bdm,bcs-bec,mag2009,unitary_review}. Calculations presented here have been performed on a larger lattice
than reported in Ref. \cite{mag2009}, namely $10^{3}$, 
with particle number varying between 90 and 110. The number of Monte Carlo samples was chosen in such a way to 
decrease the statistical errors below 1\%. The systematic errors, some due to
finite lattice effects, others due to finite range effects, are estimated at no more than 10\%,
depending on physical quantity.
In order to determine the size of the gap, we have used the one-particle temperature Green's (Matsubara) 
function which can be easily calculated within the PIMC formalism and which is related to the spectral weight function 
\cite{fw}. The PIMC method provides us with a discrete set of values of the Green's function.
In the present calculations we have determined the Green's function at $51$ imaginary time points,
equally spaced within the interval $[0, \beta = 1/T]$. We have found that due to the smooth behavior 
of the Green's function the accuracy is not affected when increasing the number of points.
We have used two methods to extract the spectral weight function from the  Matsubara function:
the maximum entropy method (MEM) and the Singular Value Decomposition method (SVD) \cite{mag2009,suppl}.

\begin{figure}[htb]
\includegraphics[width=5.4cm]{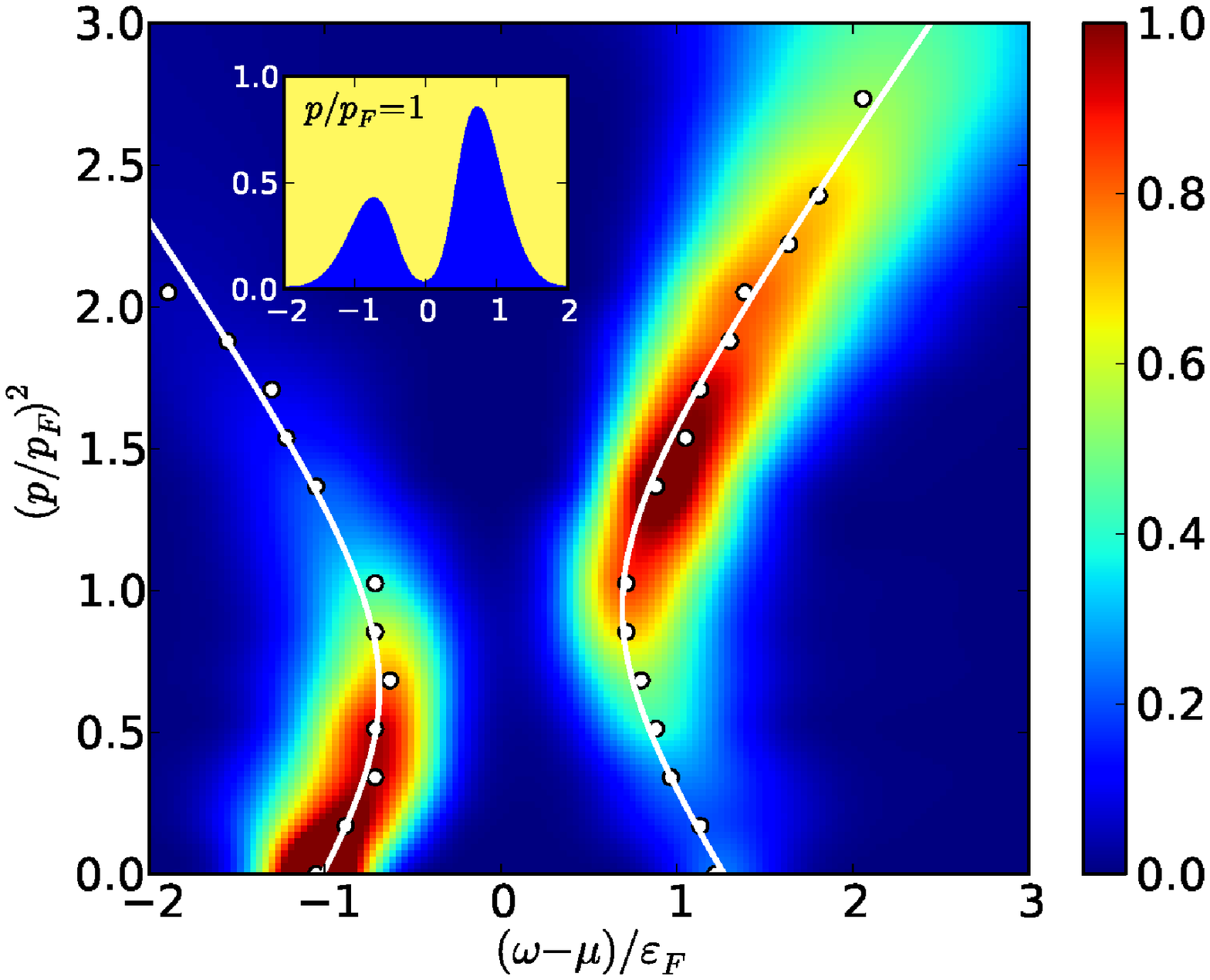}
\includegraphics[width=5.4cm]{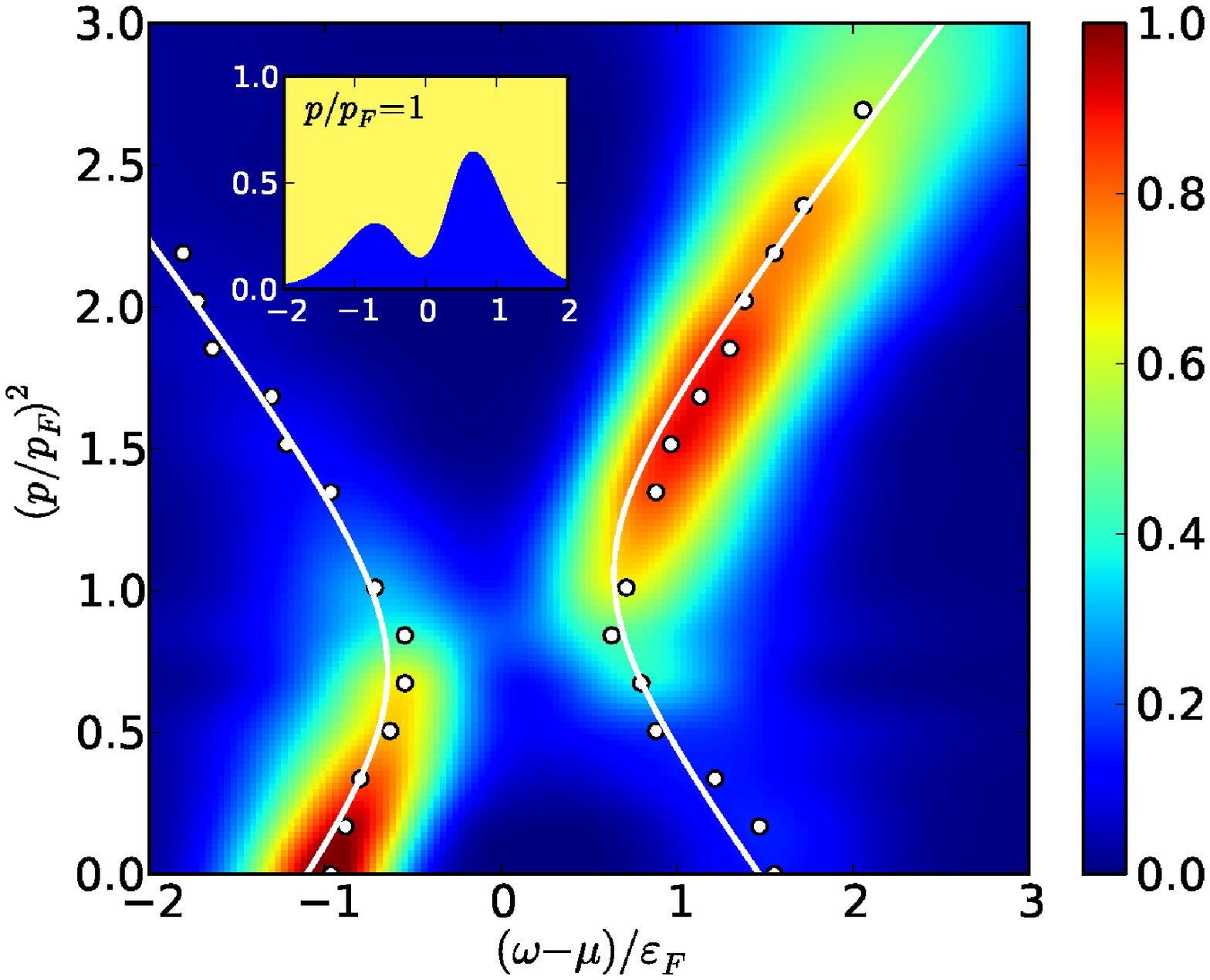}
\includegraphics[width=5.4cm]{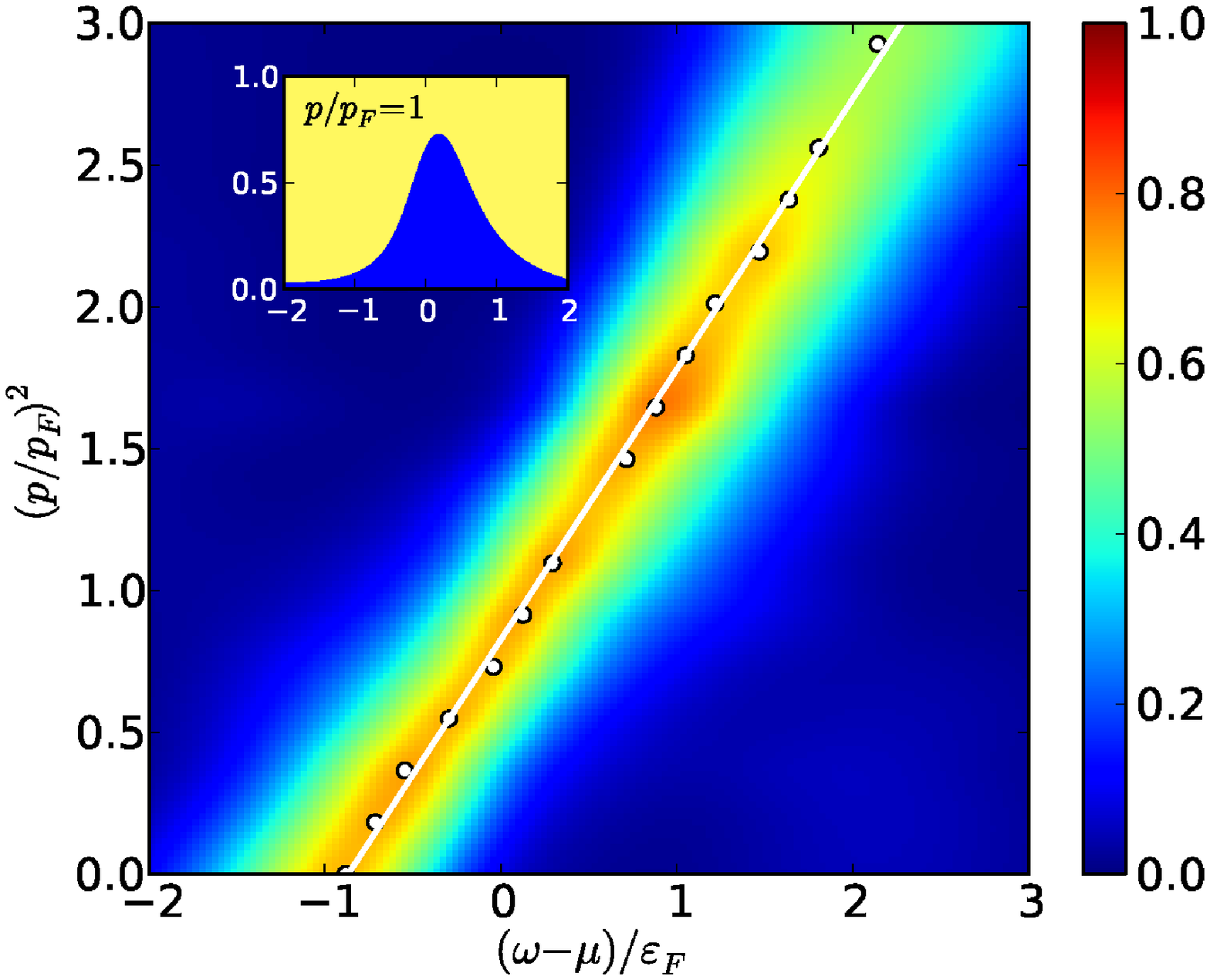}
\caption{ (Color online) Spectral weight function $A(\bm{p},\omega)$ at $1/k_{F}a = 0.2$ for three temperatures: $T=0.13\varepsilon_F < T_{c}$ (left panel), $T=0.19\varepsilon_F \approx T_{c}$ (middle panel) and $T=0.26\varepsilon_{F} > T^{*}$ (right panel). White circles denote the position of the maxima of two branches of spectral weight function and the white line represents the quasiparticle energy  $E(\bm{p})$ fit to the maxima of the spectral weight function. In the insets we show the section of the spectral weight function at the Fermi level. \label{fig:B} }
\end{figure}
\begin{table}[ht]
\begin{center}
\begin{tabular}{||c||c|c|c|c||} \hline
       & $T/\varepsilon_{F}$   & $\Delta/\varepsilon_F$    & $U/\varepsilon_F$   & $m^{*}/m$ \\
\hline
\hline
  $(k_F a)^{-1}=-0.2$                       & {\bf 0.122} & {\bf 0.21} & {\bf -0.28} & {\bf 0.95} \\
\cline{1-1} 
  $T^{*}/\varepsilon_{F} = 0.14(2)$         & {\it 0.146} & {\it 0.06} & {\it -0.44} & {\it 1.06}  \\ 
  $T_{c}/\varepsilon_{F} = 0.125(10)$       & 0.17        & 0.0  & -0.44 & 1.01  \\
\hline
\hline
  $(k_F a)^{-1}=-0.1$                       & 0.12        & 0.42 & -0.38 & 1.06   \\ 
\cline{1-1}
  $T^{*}/\varepsilon_{F} = 0.14(1)$         & {\it 0.144} & {\it 0.02} & {\it -0.31} & {\it 0.95}   \\
  $T_{c}/\varepsilon_{F} = 0.135(10)$       & 0.17        & 0.0  & -0.30 & 1.01   \\
\hline
\hline
  $(k_F a)^{-1}=-0.05$                      & 0.144 & 0.25 & -0.45 & 1.04 \\ 
                                            & 0.18  & 0.04 & -0.45 & 1.04 \\
\hline
\hline
  $(k_F a)^{-1}=0.0$                        & 0.12  & 0.47 & -0.39 & 1.0  \\
\cline{1-1}
  $T^{*}/\varepsilon_{F} = 0.19(2)$         & {\bf 0.15}   & {\bf 0.42}   & {\bf -0.38} & {\bf 1.0}  \\
  $T_{c}/\varepsilon_{F} = 0.15(1)$         & 0.165        & 0.19 & -0.48 & 1.06 \\
                                            & {\it 0.184}  & {\it 0.05} & {\it -0.47} & {\it 1.03} \\ 
                                            & {\it 0.21 }  & {\it 0.05} & {\it -0.49} & {\it 1.04} \\
\hline
\hline
  $(k_F a)^{-1}=0.1$                        & 0.14        & 0.57 & -0.37 & 0.97 \\
\cline{1-1}
  $T^{*}/\varepsilon_{F} = 0.24(2)$         & {\bf 0.17}  & {\bf 0.31} & {\bf -0.47} & {\bf 1.05} \\  
  $T_{c}/\varepsilon_{F} = 0.17(1)$         & 0.22        & 0.15 & -0.48 & 1.05 \\
                                            & {\it 0.24}  & {\it 0.0}  & {\it -0.49} & {\it 1.04} \\  
                                            & 0.28        & 0.0  & -0.51 & 1.04 \\
\hline
\hline
  $(k_F a)^{-1}=0.15$                       & 0.18  & 0.46 & -0.48 & 1.03 \\ 
                                            & 0.21  & 0.22 & -0.53 & 1.08 \\
\hline
\hline
  $(k_F a)^{-1}=0.2$                        & 0.13  & 0.66 & -0.43 & 1.0  \\
\cline{1-1}
  $T^{*}/\varepsilon_{F} = 0.26(2)$         & 0.17        & 0.62 & -0.44 & 1.01 \\ 
  $T_{c}/\varepsilon_{F} = 0.19(1)$         & {\bf 0.19}  & {\bf 0.58} & {\bf -0.43} & {\bf 1.0}  \\  
                                            & 0.22        & 0.46 & -0.45 & 1.02 \\  
                                            & {\it 0.26}  & {\it 0.03} & {\it -0.53} & {\it 1.05} \\
                                            & 0.30        & 0.0  & -0.56 & 1.05 \\
\hline
\hline
\end{tabular}
\end{center}
\caption{\label{table:results}
Pairing gaps $\Delta$, self-energy $U$ and effective masses $m^{*}$ extracted from
the spectral weight function for various temperatures $T$ and scattering lengths (see left column).
In the left column the extracted values of crossover temperature $T^{*}$ have been shown
together with the values of the critical temperatures (upper estimates) taken from Ref. \cite{bcs-bec}.
All the results except of those for $(k_{F}a)^{-1}=-0.2$ have been obtained
for the lattice $10^3$. The results on the BCS side at $(k_{F}a)^{-1}=-0.2$
have been obtained for the lattice $8^{3}$ as described in \cite{mag2009}.
The values of
$\Delta/\varepsilon_F < 0.1$ should be treated as lying within the interval $[0,0.1]$ due 
the finite resolution, see Ref. \cite{suppl}.}
\end{table}
\begin{figure}
\includegraphics[width=6.5cm]{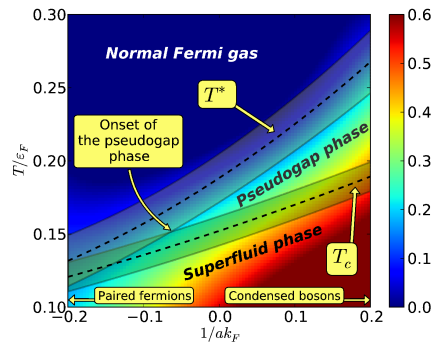}
\caption{ (Color online) The gap $\Delta/\varepsilon_F$ extracted from the spectral weight function as a function of temperature and scattering length. The dashed lines denote two temperatures: critical temperature $T_c$ and the crossover temperature $T^{*}$. Uncertainties (both systematic and statistic errors, estimated to be no more than 10\%) of these temperatures, are denoted by shaded area.  
\label{fig:C} }
\end{figure}
\begin{figure}
\includegraphics[width=6.0cm]{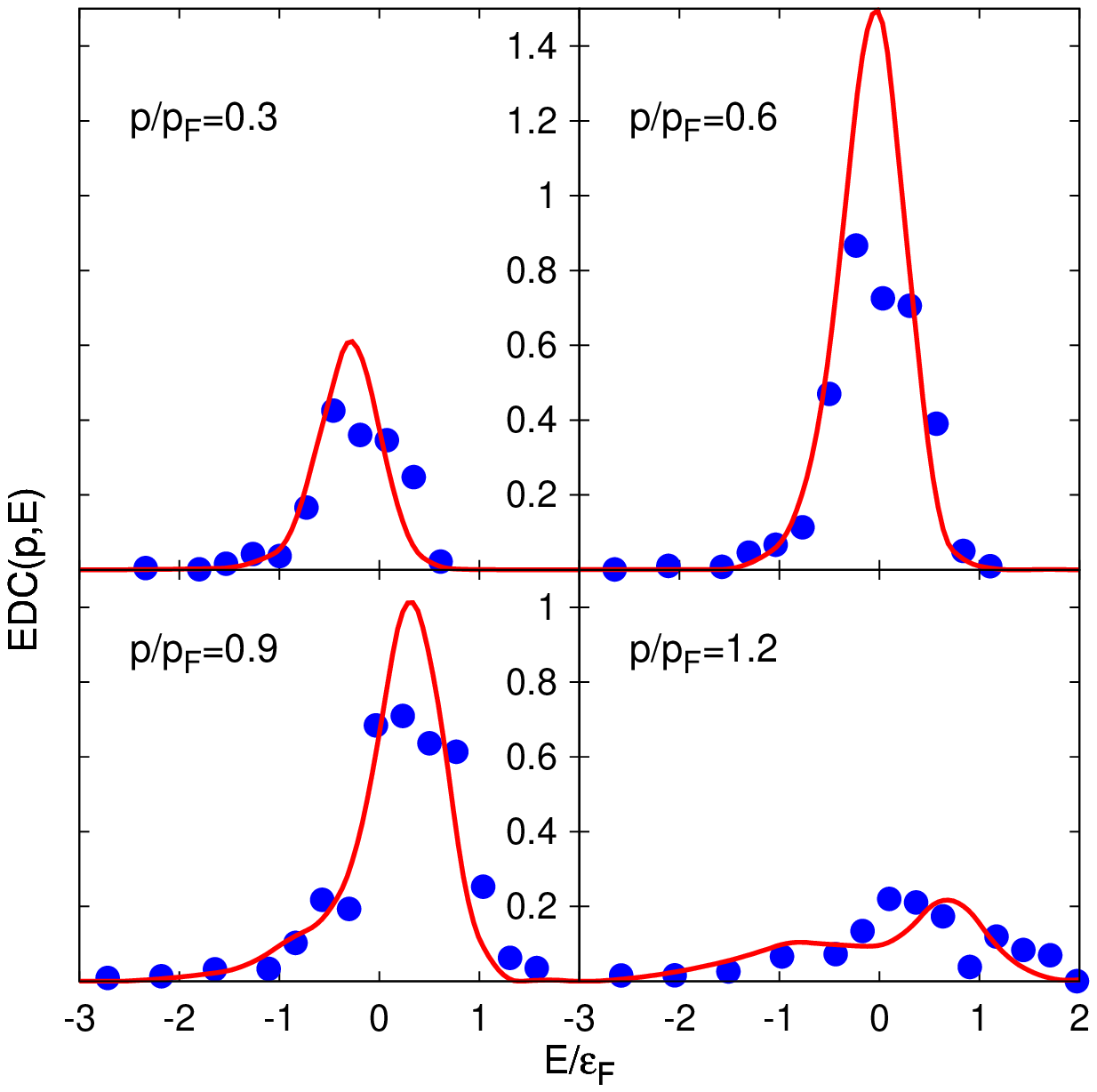}
\caption{ (Color online)
Experimental (dots) and theoretical (lines) energy distribution curves (EDC)
for the trapped atomic gas in the unitary regime at the critical temperature
(defined in the center of the trap) for various momenta $p$.
\label{fig:D} }
\end{figure}

As an example in the Fig. \ref{fig:B} the spectral weight function is shown on the BEC side of the unitary regime for $(k_{F}a)^{-1}=0.2$. The three temperature values correspond to three regimes realized in this system. At $T\approx 0.13\varepsilon_{F}<T_{c}=0.19(1)\varepsilon_F$ the system is superfluid  \cite{bcs-bec}, at $T\approx 0.19\varepsilon_{F}$ the long range order is lost, but the gap in the single particle spectrum is still present. Finally, the lower panel shows the spectral weight function just above the crossover to the normal Fermi liquid. The negative energy branch is nonzero essentially up to $p\approx p_{F}$. For the momenta above the Fermi surface the negative energy branch has a tail which in the case of the unitary Fermi gas is related to the behavior of the occupation probability $n(p)$ at large $p$ which decay as $1/p^{4}$ \cite{schneider}. The positive energy branch starts essentially around $p\approx p_{F}$ and the presence of the gap can be easily detected. The upper panel of Fig. \ref{fig:B} describes the superfluid system with no states at the Fermi surface (see inset). In the middle panel ($T\approx T_c$) the density of states at the Fermi level is significantly lowered. At large temperatures the two branches merge into a single branch extending from negative to positive values of $\omega - \mu$, characteristic for a normal state. Similarly like in \cite{slda,mag2009} we have fitted the formula for the quasiparticle energy $E(\bm{p}) = \sqrt{ \left ( p^2/2m^{*}+U-\mu \right )^2+\Delta^2}$ to the maxima of the spectral weight function. 
The extracted values of gap $\Delta$, self-energy $U$ and effective mass $m^*$ have been presented in the table \ref{table:results}.
One may notice that the effective mass is practically constant and does not deviate from the bare mass by more than $10$\%. 
The self-energy $U$ decreases with $(k_{F}a)^{-1}$. When compared to the experimentally extracted value at unitarity 
(see Ref. \cite{schirotzek2008}) it agrees within a $10$\% accuracy. The values of $T_{c}$ around unitarity have been determined 
elsewhere \cite{bcs-bec,burovski}. The value of the gap at $T_{c}$ is usually called the 
pseudogap, i.e. measure of the strength of the short range correlations at the critical temperature. These results can 
be subsequently used to estimate the value of $T^{*}$, where the gap tends to zero and also the value of the gap at 
the critical temperature $T_{c}$ for the superfluid to normal phase transition. 
We show the size of the gap $\Delta/\varepsilon_F$ plotted as a function both temperature and scattering length in Fig. \ref{fig:C}. The two temperatures $T^{*}$ and $T_{c}$ become indistinguishable, due to uncertainties generated by Monte Carlo approach, between $-0.1 < (k_{F}a)^{-1} < 0$. Hence already at the unitary limit the system exhibits the pseudogap phase, albeit in a relatively small temperature interval. The temperature interval within which the pseudogap phase exists increases as one moves to the BEC side. 

The present work is the first attempt to determine the evolution of the pseudogap as a function of temperature and 
interaction strength from fully {\it ab initio} calculations. Previous results obtained using this {\it ab initio } 
method were used as a benchmark by other theoretical approaches and were as well validated by subsequent experimental 
results \cite{luo,salomon1}. 
On one hand, the most recent calculations of Ref. \cite{perali}, based on the pairing-fluctuation theory, 
indicated the presence of the pseudogap without however specifying its size and evolution as a function of 
the scattering length. On the other hand the recent measurements of the magnetic susceptibility of the trapped 
atomic gases were explained within the Fermi liquid theory without invoking the pseudogap model \cite{salomon1,salomon2}, a description challenged in Ref. \cite{perali}. The Fermi liquid parameters in Ref. \cite{salomon2} were 
extracted in a QMC calculation at zero temperature, in which the formation of pairs was artificially suppressed, 
and the spectral functions were approximated with $\delta$-functions. Here we present the comparison of the experimental 
results based on the wave-vector resolved spectroscopy \cite{gaebler2010,perali,hightc_apres}. Since the experimental data are for 
a trap which is inhomogeneous and likely composed of spatially separated and coexisting phases it is difficult to 
relate the particular behavior of the extracted energy distribution curves to the appearance of the pseudogap phase. 
However the agreement of the experimental and theoretical results supporting the existence of the pseudogap provides 
an implicit confirmation of the quality of the theoretical calculations.  We use the Local Density Approximation \cite{bdm2} 
to relate the PIMC input to trap data, see Fig. \ref{fig:D} (without using any fitting parameters, see \cite{suppl} for details). 
Over the years there 
was a significant body of work addressing the physics of preformed pairs, pseudogap and dissociation temperature $T^*$, 
see Refs.  \cite{randeria,levin,strinati,hui,mohit} and earlier references, based on a variety of theoretical models 
and approximations. The theoretical predictions for some or all of these entities either varied widely or were at most 
of qualitative nature and the values obtained in these references for the pseudogap and $T^*$ are significantly different 
from ours. The lack of quantifiable error bars makes essentially impossible to re-conciliate these earlier predictions 
and to judge their accuracy.
 
In summary we have determined in {\it ab initio} calculations the spectral weight function for various temperatures 
around the unitary regime. We have confirmed the existence of the pseudogap phase (we have reported the evidence 
of this phase at unitarity in Ref. \cite{mag2009}) and located the onset of this phase at a value of the interaction 
strength corresponding to $(k_{F}a)^{-1}\approx -0.05$. To date the present results are the first {\it ab initio} 
calculations which were able to determine the size of the pseudogap and its behavior as a function of interaction 
strength and temperature. Our results indicate that a finite temperature structure of the Fermi gas around unitarity 
is more complicated and involves the presence of the phase with preformed Cooper pairs, which however do not contribute 
to the long range order. 


\begin{center}
{\bf Acknowledgments}
\end{center}
We thank T.E. Drake for providing the experimental data \cite{gaebler2010}. Support is acknowledged from the DOE under grants DE-FG02-97ER41014 and DE-FC02-07ER41457, from the Polish Ministry of Science under contract N N202 128439. Calculations were performed on the UW Athena cluster and at the Interdisciplinary Centre for Mathematical and Computational Modelling (ICM) at Warsaw University. 


\newpage

\begin{center}
{\em \large Supplemental online material for} \\
{\bf \large The onset of the pseudogap regime in ultracold Fermi gases}
\end{center}

\bigskip

\section{ Spectral weight function and single particle parameters} 

To extract the spectral weight function we have used
the one-body temperature Green's (Matsubara) function
which can be calculated within the PIMC formalism \cite{fw}:
\beq
{\cal G }(\bm{p},\tau)=
-\frac{1}{Z} \tr \{\exp[-(\beta-\tau) (\hat{H}-\mu \hat{N})]\hat{\psi}(\bm{p})\exp[-\tau(\hat{H}-\mu \hat{N})]\hat{\psi}^\dagger(\bm{p}) \}. \label{eq:Gp}
\eeq
In the above expression $\beta = 1/T$ is the inverse temperature and $0 \leq \tau \leq \beta$. 
For the spin-symmetric system with the spin-independent interaction
${\cal G }(\bm{p},\tau)$ is diagonal in the spin space
and contributions coming from the spin-up and spin-down fermions are equal.
Therefore we have suppressed the spin indices in all formulas.
The trace $\tr$ is
performed over the Fock space, and $Z=\tr \{\exp[-\beta(\hat{H}-\mu \hat{N})]\}$. The spectral weight function $A(\bm{p},\omega) = A^{(+)}(\bm{p},\omega) + A^{(-)}(\bm{p},\omega)$,
is defined as:
\bea
A^{(+)}(\bm{p},\omega) & = & \frac{2\pi}{Z}\sum_{n,m}\exp[-\beta(E_{n}-\mu N_{n})]\langle n |\hat{\psi}(\bm{p})| m \rangle\langle m |\hat{\psi}^\dagger(\bm{p})| n \rangle
                              \delta(\omega + E_{n} - E_{m} + \mu), \nonumber \\
A^{(-)}(\bm{p},\omega) & = & \frac{2\pi}{Z}\sum_{n,m} \exp[-\beta(E_{n}-\mu N_{n})]\langle n |\hat{\psi}^\dagger(\bm{p})| m \rangle\langle m |\hat{\psi}(\bm{p})| n \rangle
                             \delta(\omega + E_{m} - E_{n} + \mu),
\eea
where $| n \rangle$ represents the state with energy $E_{n}$ and particle number $N_{n}$. It
is related to the temperature Green's function:
\beq
{\cal G }(\bm{p},\tau)=-\frac{1}{2\pi}\int_{-\infty}^{\infty}
d\omega A(\bm{p},\omega)\frac{\exp(-\omega\tau)}{1+\exp(-\omega\beta)},
\label{eq:Ap}
\eeq
and by definition $A(\bm{p},\omega)$ fulfills the following constraints:
\beq
A(\bm{p},\omega) \ge 0, \quad \quad
\int_{-\infty}^{\infty}\frac{d\omega}{2\pi} A(\bm{p},\omega) = 1, \quad
\int_{-\infty}^{\infty}\frac{d\omega}{2\pi} A(\bm{p},\omega) \frac{1}{1+\exp(\omega\beta)}=  n(\bm{p}), 
\label{eq:Ap_con}
\eeq
where $n(\bm{p})$ is the occupation number of a state with momentum $\bm{p}$.
The PIMC method provides us with a discrete set of values of the Green's function.
In the present calculations we have determined the Green's function at $51$ imaginary time steps,
equally spaced within the interval $[0, \beta]$. We have found that due to the smooth behavior 
of the Green's function the accuracy is not affected when increasing the number of points.
The discretization of the left hand side of Eq.~(\ref{eq:Ap}) leads to the class of 
numerically ill-posed inverse problems. 
We have used two methods, designed particularly to deal with such problems:
the maximum entropy method (MEM) and the Singular Value Decomposition method (SVD).

\begin{figure}[htb]
\includegraphics[width=7.7cm]{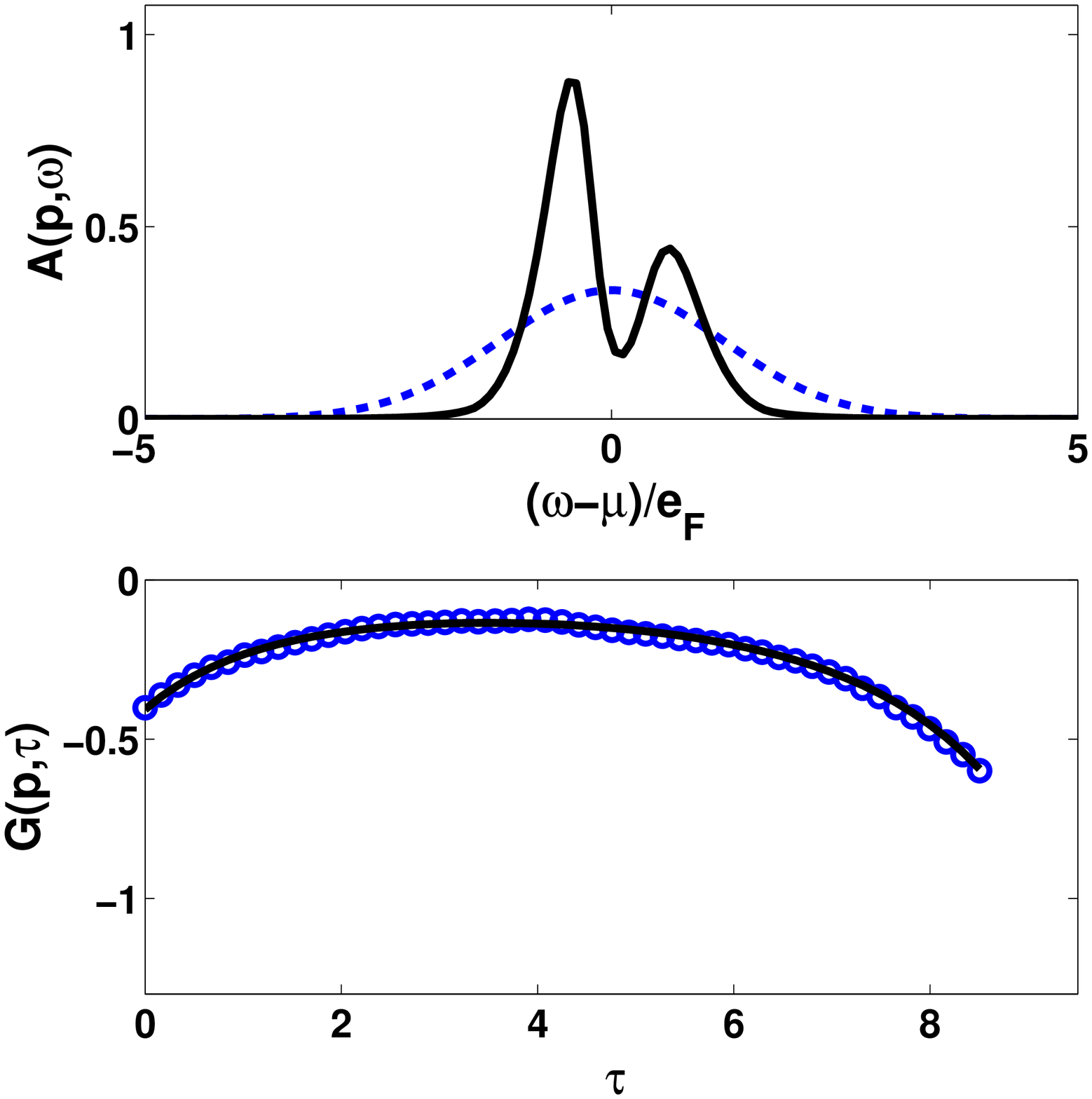}
\includegraphics[width=7.7cm]{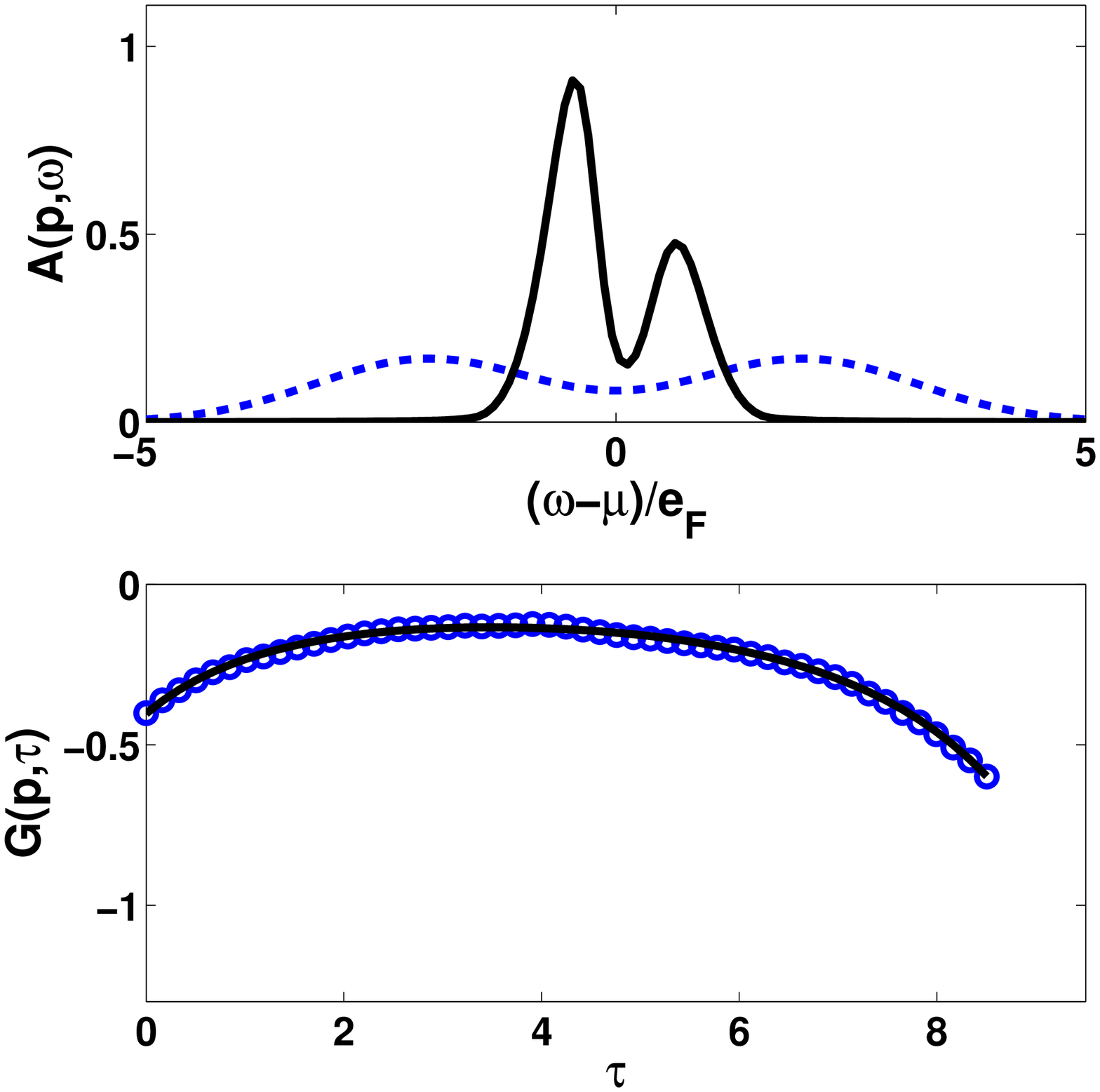}
\caption{ (Color online)
Upper subfigures: The extracted spectral weight function (line) at the unitary regime for the
momentum at the vicinity of the Fermi surface (occupation probability
$n(p)\approx 0.6$) at $T/\varepsilon_{F} = 0.12$ ($T_{c}/\varepsilon_{F} = 0.15(1)$). Two sets of results have been obtained
using MEM for two different models (dashed line) being one gaussian (left figure)
and two gaussians (right figure). The lower subfigures show the Green's function
(open circles) obtained from PIMC calculations and Green's function reconstructed from the
spectral weight function (line) presented in the upper subfigures.
\label{apomega} }
\end{figure}

In the maximum entropy method \cite{jaynes}, which is based on Bayes' theorem,
we treat the values of the Green's function calculated within PIMC approach 
as normally distributed random numbers ${\cal \tilde{G}}(\bm{p},\tau_{i})$ ($i=0,1,2,...,{\cal
N}_{\tau}=50$) around the true values ${\cal G}(\bm{p},\tau_{i})$. The  Bayesian strategy
of searching for the most probable solution leads to the minimization
of the quantity $\frac{1}{2}\chi^{2} - \alpha S({\cal M})$, where $\alpha>0$,
\beq
\chi^{2}=\sum_{i=1}^{{\cal N}_{\tau}}
\left [{\cal \tilde{G}}(\bm{p},\tau_{i}) - {\cal G}(\bm{p},\tau_{i})
\right]^{2}/ \sigma^{2},
\eeq 
and $S({\cal M})$ is the relative information entropy with
respect to the assumed model ${\cal M}$:
\beq
S({\cal M})= -\sum_{k}\Delta\omega \biggl [ A(\bm{p},\omega_{k}) - {\cal
M}(\omega_{k}) - A(\bm{p},\omega_{k})\ln
\left ( \frac{A(\bm{p},\omega_{k})}{{\cal M}(\omega_{k})} \right ) \biggr ]
. \label{eq:method}
\eeq
The minimization is performed with respect to the values of $A(\bm{p},\omega_{k})$.
The entropy term prevents excessive inclusion of unjustified structure into
the shape of the spectral weight function. The constraints (\ref{eq:Ap_con})
are enforced by means of Lagrange multipliers.
In our previous calculations obtained for the smaller lattice of size 
$8^3$, which were reported in Ref. \cite{mag2009}, the calculated Green's function 
was not as smooth as the one we obtained for $10^3$ lattice. Therefore in the present
calculations we could decrease the value of  $\alpha$ by an order of magnitude
as compared to the previously used value: $\sigma^{2}\alpha \approx 0.3$.
It has substantially improved the resolution of the present calculations.
We have used the energy window $-5 < (\omega-\mu)/\varepsilon_{F} < 5$
and calculated $A(\bm{p},\omega_{k})$ for $120$ points equally spaced within this interval, although
the size of the window and the number of points could have been chosen smaller without any effect
on the results.
In our MEM approach we have improved the method of finding the spectral weight function
by constructing a sequence of minimizations with a gradually refined model. Namely, at the end
of each minimization process the result has been used to defined a new model in the form 
of two gaussians which maximize the overlap: 
$\sum_{k} \sqrt{ A(\bm{p}, \omega_{k}){\cal M}(\omega_{k}) }\Delta\omega/2\pi$. Note that since the spectral
weight function and the model are both nonnegative and normalized, the above quantity can
take values from the interval $[0,1]$ only.
The parameters of the model ${\cal M}$ include the heights, widths and the relative shift of the gaussians.
These parameters were adjusted to maximize the overlap and thus define a new model.
Then the new minimization process was started. The procedure has been
continued until the value of $\chi^2$ did not improve.
The procedure has generated spectral weight functions, where
both the position of the maxima and the widths were surprisingly
stable, no matter what particular model has been used to begin the minimization process (see Fig. \ref{apomega} ).

The Fig. \ref{apomega} presents the spectral weight function at the unitary regime,
for the momentum close to the Fermi surface, at the temperature which is slightly
smaller than the critical value: $T/\varepsilon_{F}=0.12$. This example
represents the most difficult situation (from the numerical point of view), where the weight function 
exhibits the well defined structure of two maxima around the Fermi level. 
However, as one can see, the calculations are surprisingly
stable and starting from two completely different models one gets practically the same
results. It has to be emphasized that the results exhibit also the
stability of the widths of the spectral weight function.

The second approach is based on the singular value decomposition of
the integral kernel (\ref{eq:Ap}) \cite{svd1}. It was shortly described in our previous paper \cite{mag2009}. 
Here we mention that the numerical realization of the SVD approach requires to introduce two parameters 
($a$ and $b$) which define the interval ($a < \omega < b$), where $A(\bm{p},\omega)$ is assumed to be localized:
\beq
\int_{-\infty}^{\infty} d\omega A(\bm{p},\omega)\frac{\exp(-\omega\tau)}{1+\exp(-\omega\beta)}
= \int_{a}^{b} d\omega A(\bm{p},\omega)\frac{\exp(-\omega\tau)}{1+\exp(-\omega\beta)}.
\eeq
The prior information
concerning the support of the spectral weight function is essential to produce the results of high quality \cite{svd2}. 
Since the optimal values of parameters $a$ and $b$ are unknown we have applied the so called \textit{bootstrap} strategy~\cite{bootstrap}. 
The procedure consists of many reconstructions of the spectral weight function
performed for randomly generated parameters $a$ and $b$ (within some reasonably chosen interval). 
These partial solutions are subsequently used to evaluate 
the final solution (as an average) as well as uncertainties with respect to algorithm parameters. 

\begin{figure}[htb]
\includegraphics[width=9.3cm]{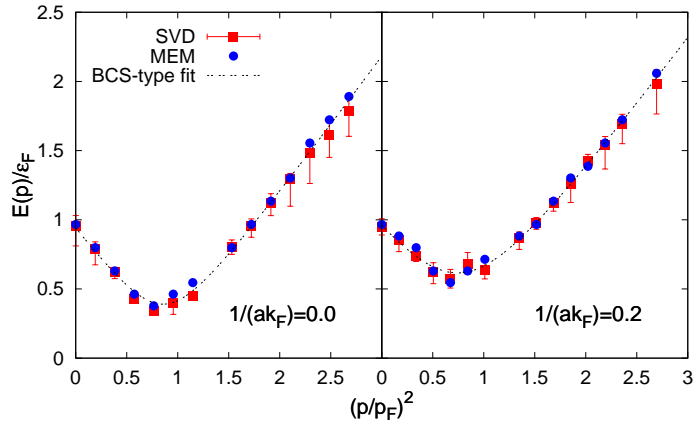}
\caption{ (Color online) Quasiparticle energies (maxima of the spetral weight function at fixed momentum)
obtained using MEM and SVD methods for the unitary limit (left panel) and the BEC side 
(right panel) at the critical temperature. The error bars for the SVD solutions
are related to the sensitivity of the maxima localization with respect to the parameters of the SVD algorithm and 
Monte Carlo uncertainties. In both cases the presence of the gap is revealed. 
Dashed line denotes the fit of the BCS dispersion relation to the extracted quasiparticle energies.
\label{crosscheck} }
\end{figure}

Since MEM and SVD methods are based on completely different approaches
their agreement serves as a robust test for the correctness of the determination of the spectral weight function.
In the Fig.~\ref{crosscheck} the quasiparticle energies obtained using MEM and SVD methods have been presented and
the quantitative agreement between both methods is clearly visible.

We have performed a number of tests to quantify the limitations associated with MEM and SVD approaches. 
In both cases we have identified the \textit{resolution limit} which impose the lower bounds on the value of the gap
which can be detected by both methods. In the case of cold atomic gases in the unitary regime the resolution limit  
is about $0.19\varepsilon_{F}$ which turns out to be sufficient to reveal the existence of the pseudogap phase.
It means that both methods provide in practice only a lower bound on the temperature $T^{*}$.

Similarly like in Ref. \cite{mag2009} we have fitted the formula for the quasiparticle energy $E(\bm{p})$ to the maxima of the spectral weight function. The extracted values of gap $\Delta$, self-energy $U$ and effective mass $m^*$ have been presented in the table I. The comparison of the results provided in the Table I and those obtained in Ref. \cite{mag2009} for a smaller lattice
allows to estimate errors related to the finite size effects. The single-particle parameters do not vary more than by 10\%,
except for the gap at the temperatures close to $T^{*}$, where it is on the verge or below the resolution limit.
The resulting uncertainty of the value of $T^{*}$ have been included in the error bars presented in the Fig. 3
in the article.  

\section{ Energy distribution curves from the spectral weight function} 

Since the experiments with cold atoms are performed in a trap one has to translate
the results obtained for the uniform system to the nonuniform one determined by the geometry of a trap.
The quantity of interest is the energy distribution curve (EDC), which is measured
by exciting atoms from strongly interacting to weakly interacting states.
Hence the results depends on the occupation of the interacting state and
thus are related to the spectral weight function:
\beq
\EDC(p, E, T)={\cal C}p^{2}
             \int_{0}^{\infty} dr\,r^{2} \frac{1}{\varepsilon_{F}(r)}
             A\Big[ \frac{p}{p_{F}(r)},\frac{E-\mu(r)}{\varepsilon_{F}(r)}, \frac{T}{\varepsilon_{F}(r)} \Big]
             f(E-\mu(r)),
\label{edc}
\eeq
where $k_{F}(r), \mu(r), \varepsilon_{F}(r)$ are the local Fermi momentum, chemical potential and Fermi energy in the trap, respectively.
Here the spectral weight function is expressed in units $[\varepsilon_{F}^{-1}]$ and parameter ${\cal C}$ is chosen to make the total area of the $\EDC$ signal equal to unity, and $f(E)=1/( 1 + e^{\beta E} )$ is the Fermi distribution function at the temperature $T=1/\beta$ which acts as a filter for the occupied branch of the spectral function.

In order to calculate the above expression one has to determine the profile of atomic density in a trap. 
This has been done using the local density approximation (LDA) \cite{bdm2}.
In this approach, the grand canonical thermodynamic potential for a system confined by a trap potential 
$U(\bf{r})$ is a functional of the local density $n(\bf r)$ given by
\begin{equation}
\Omega = \int dV \left [
\frac{3}{5}\veps_F({\bf r})\varphi (x, k_{F}({\bf r})a)n({\bf r}) +
U({\bf r}) n({\bf r})-\lambda n({\bf r}) \right ],
\end{equation}
where 
\begin{equation}
\label{eq:XandEF}
x = \frac{T}{\veps_F({\bf r})}, \quad 
\veps_F({\bf r}) = \frac{\hbar^2}{2m} [3 \pi^2  n({\bf r})]^{2/3}, \quad
\frac{F}{N}=\frac{3}{5}\veps_F\varphi(x, k_{F}({\bf r})a),
\end{equation}
where  $F/N$ is the free energy per particle. 
The overall chemical potential $\lambda$ and the temperature $T$ are constant throughout the system. 
The density profile will depend on the shape of the trap as dictated by $\delta \Omega / \delta n({\bf r}) = 0$, which results in: 
\begin{equation}
\label{eq:chempot}
\frac{\delta \Omega}{\delta n({\bf r})}=
\frac{\delta(F-\lambda N)}{\delta n({\bf r})}
=\mu(x({\bf r}), k_{F}({\bf r})a) + U({\bf r})-\lambda .
\end{equation}
At a given $T$ and $\lambda$, equations (\ref{eq:XandEF}) and (\ref{eq:chempot}) completely determine the density profile $n(\bf r)$
in a given trap for a given total particle number.

\begin{figure}[htb]
\includegraphics[width=7.3cm]{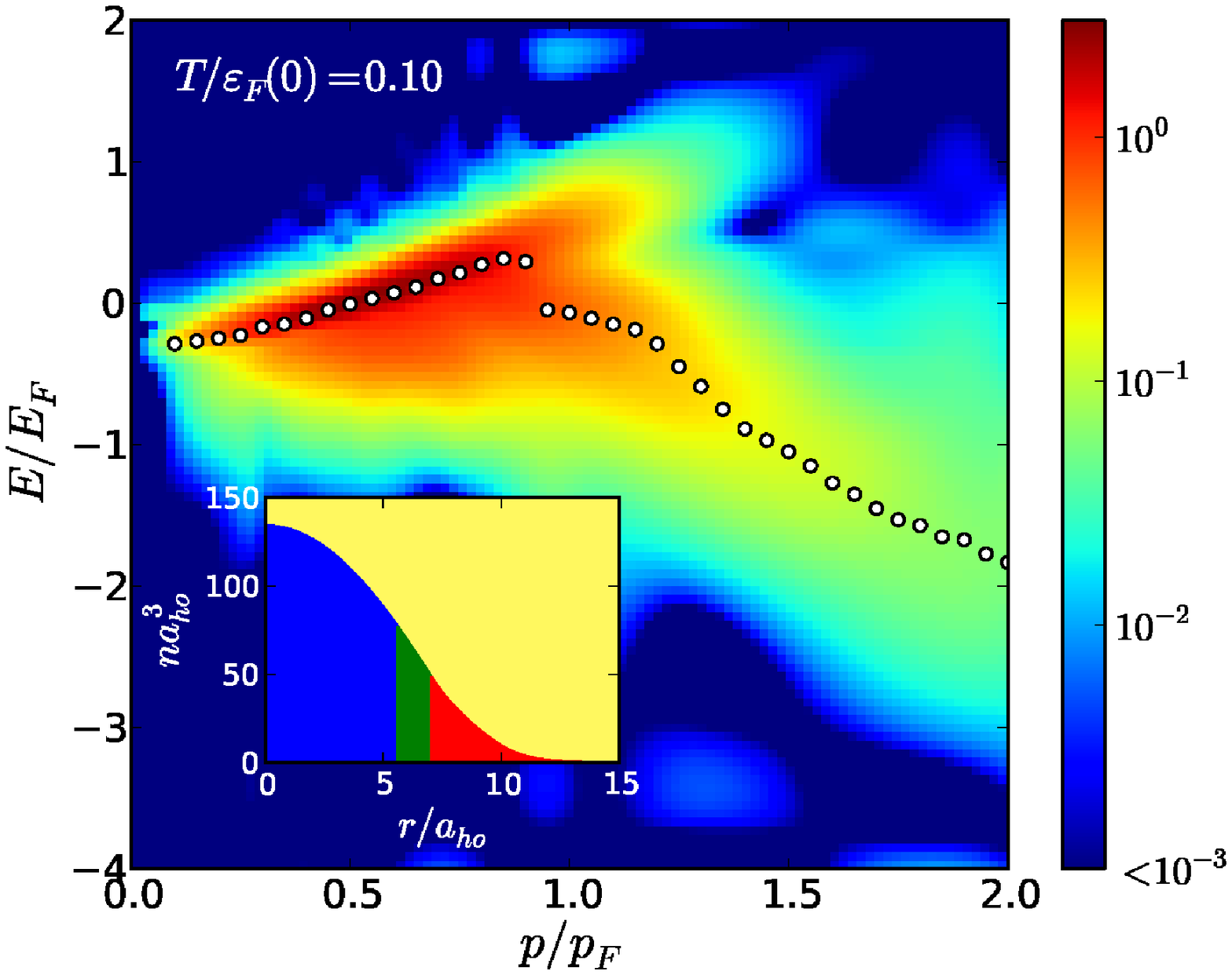}
\includegraphics[width=7.3cm]{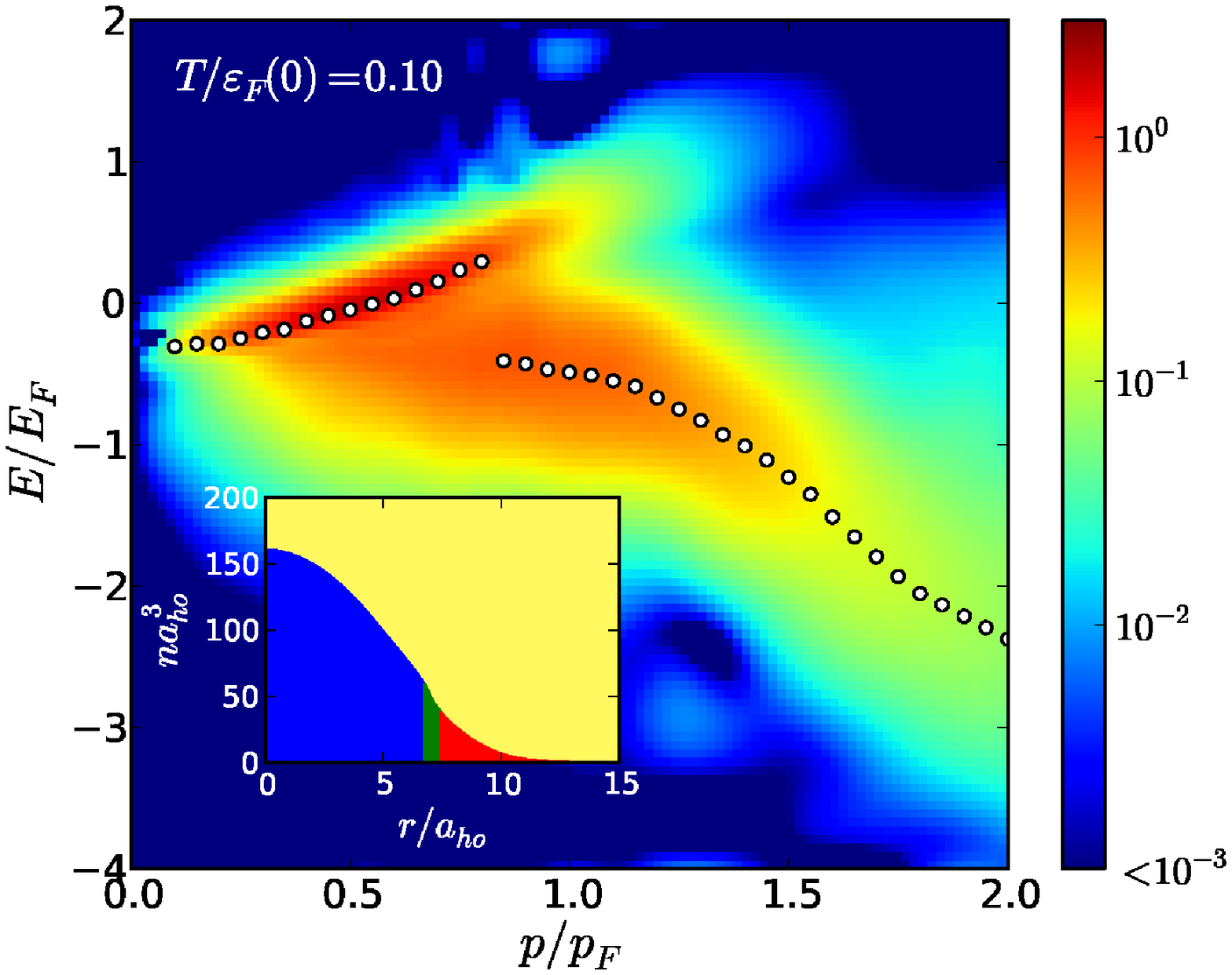}
\includegraphics[width=7.3cm]{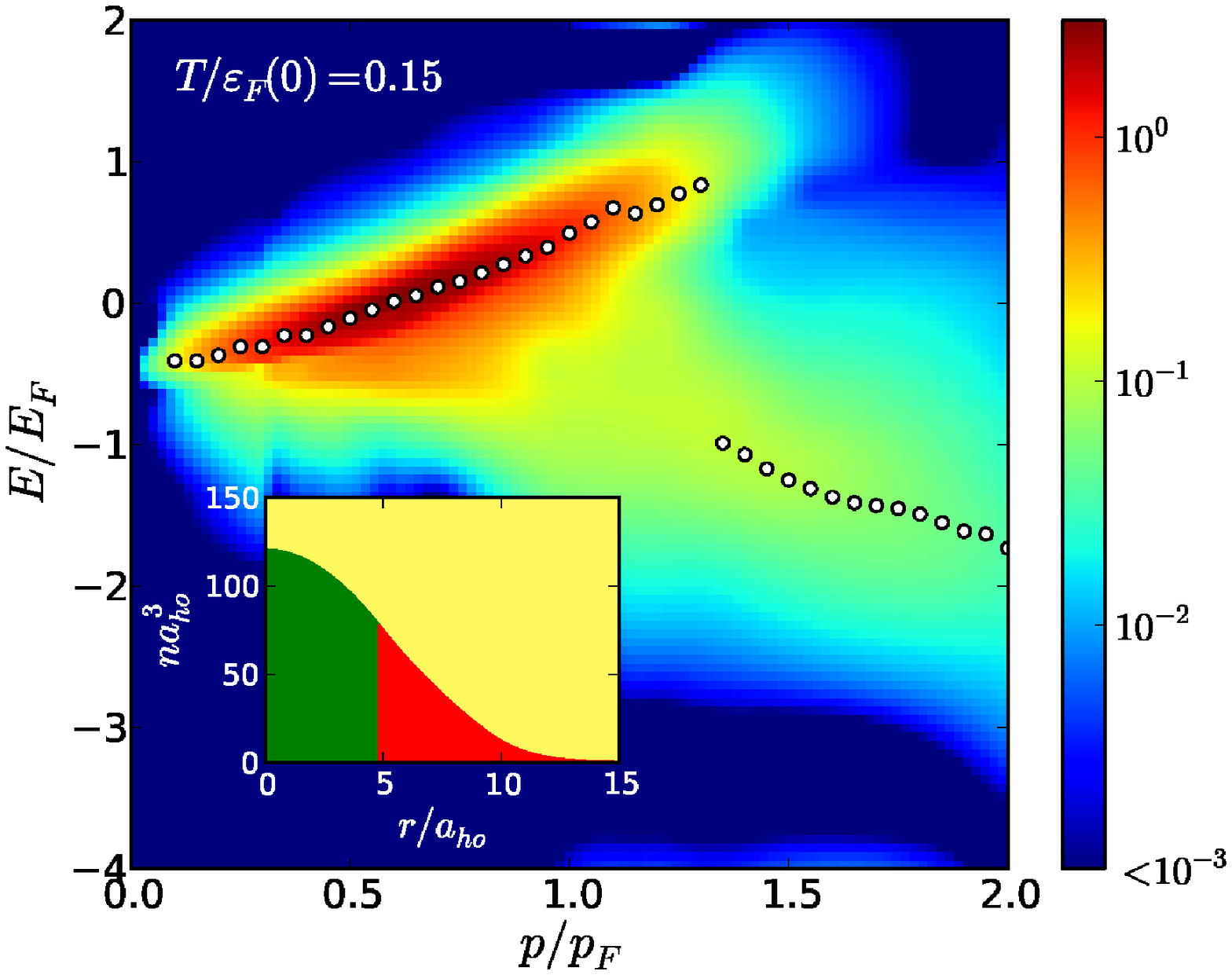}
\includegraphics[width=7.3cm]{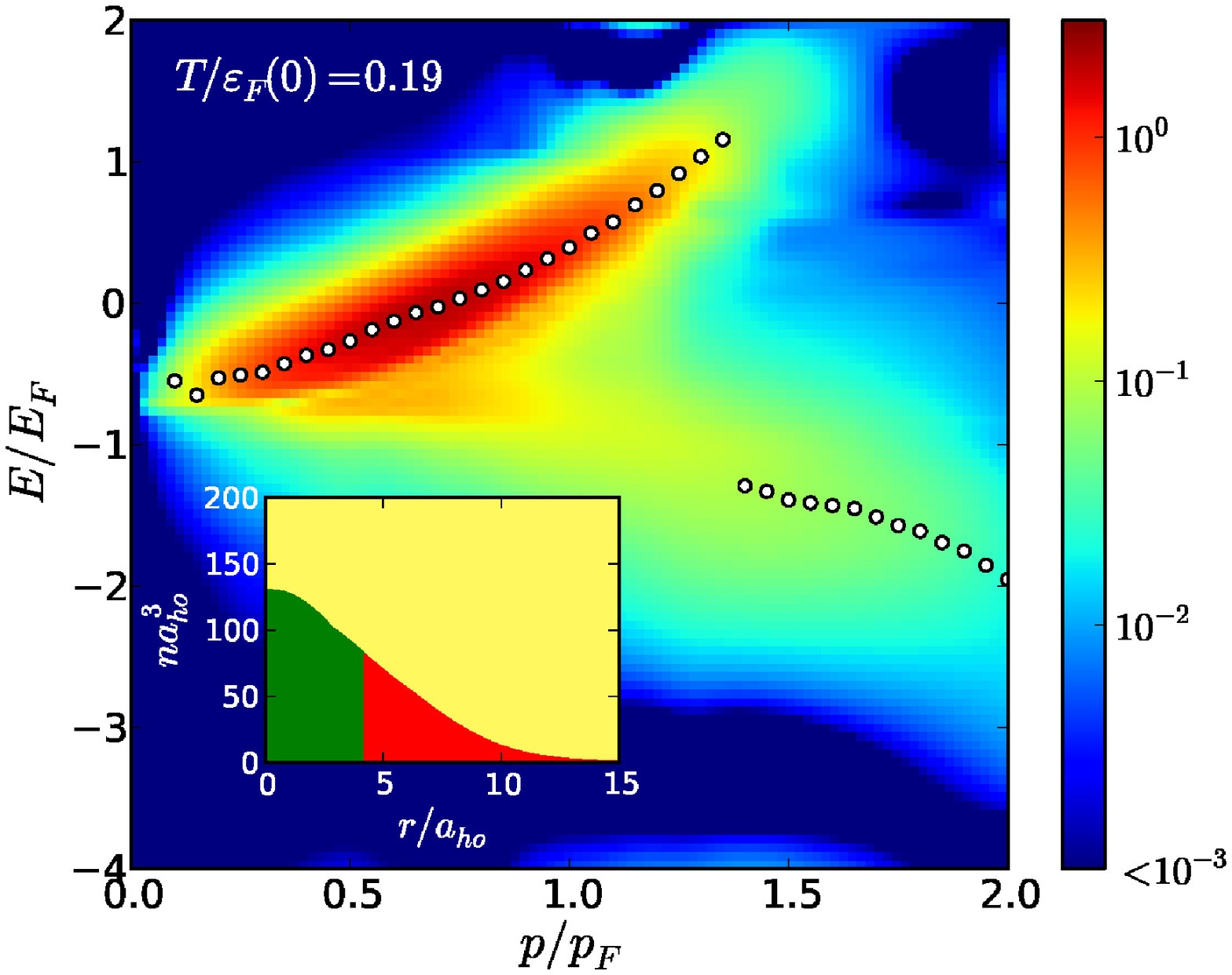}
\includegraphics[width=7.3cm]{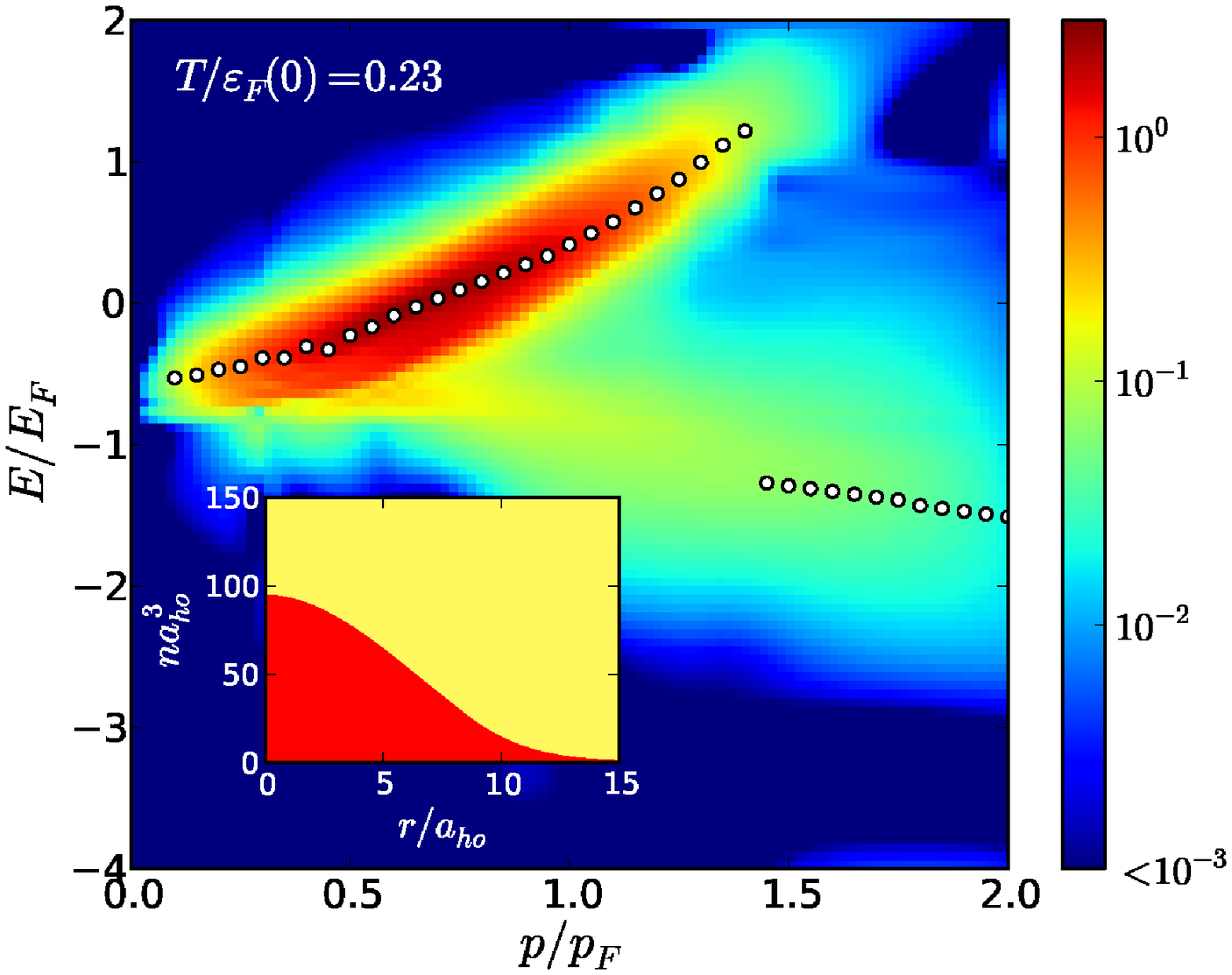}
\includegraphics[width=7.3cm]{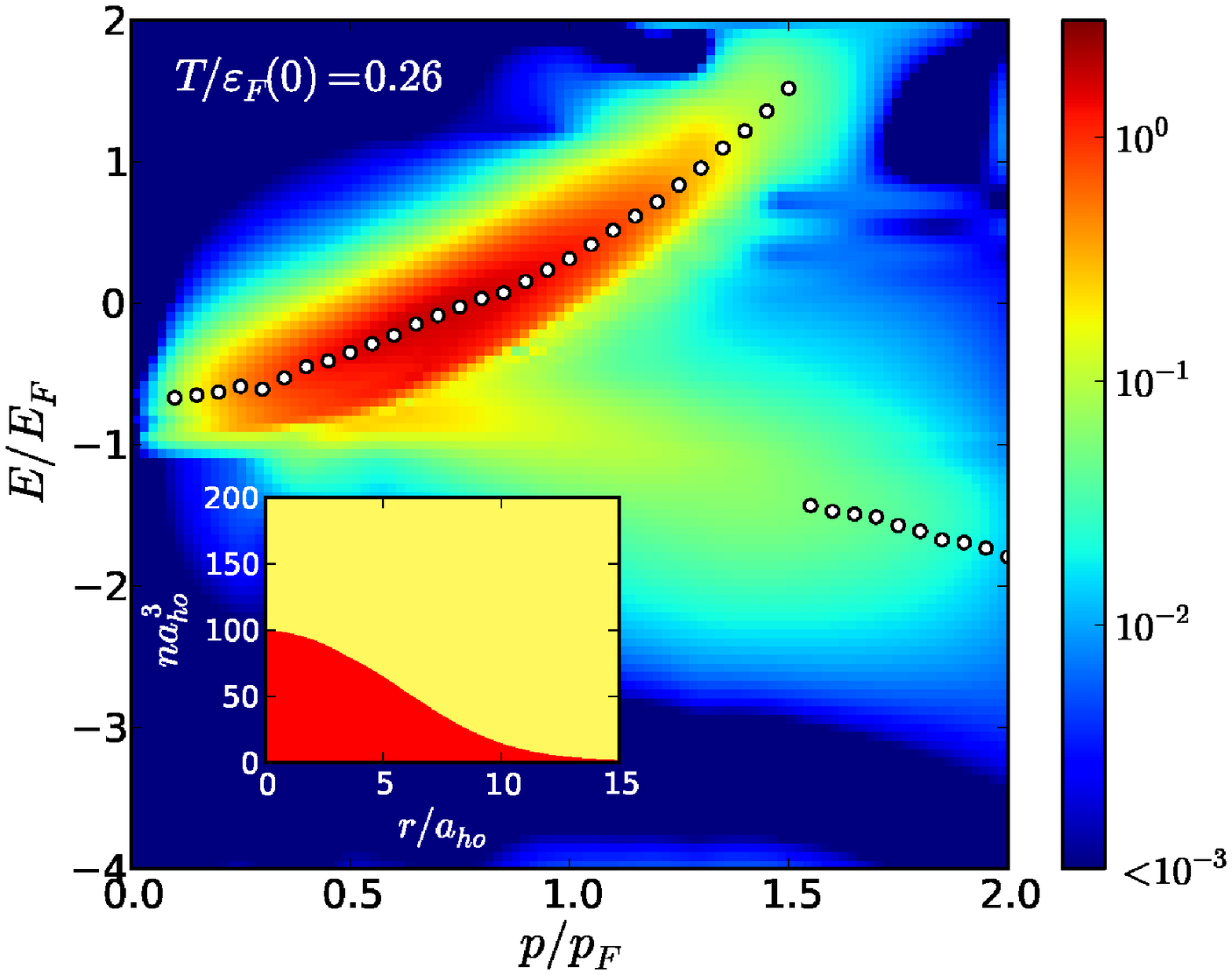}
\caption{ (Color online)
Energy distribution curves obtained from PIMC calculations in the LDA approximation
for three temperatures (defined in the center of the trap) for JILA trap \cite{gaebler2010}. 
The left column corresponds to the unitary regime, and the right one - to the BEC side
corresponding to $(k_{F} a)^{-1}=0.2$ defined at the center of the trap. $E_{F}=\hbar\omega_{0} (2N)^{1/3}$ is 
the Fermi energy of the gas in the harmonic 
trap defined by the frequency $\omega_{0}= ( \omega_{x}\omega_{y}\omega_{z} )^{1/3}$.
In the middle subfigures the temperature corresponds to the critical temperature.
The insets show the density distribution sections. Different colors correspond
to different phases: superfluid phase - blue, pseudogap phase - green, normal phase - red.
\label{edc_fig} }
\end{figure}

For a given profile of the atomic cloud the integral (\ref{edc}) can be determined. 
The results presented here concerns the EDC in the vicinity of the
critical temperature $T_{c}$. 
As an example we show 2D plots of the EDC at unitarity and at the BEC side corresponding
to $(k_{F} a)^{-1}=0.2$ at the center of the trap (see Fig. \ref{edc_fig}). Results were generated for the trap 
used in the experiment reported in Ref. \cite{gaebler2010}.
If the system is normal the maxima of the EDC should reproduce the $E\propto p^{2}$ dependence of the dispersion relation.
At the lower temperatures, where part of the system is superfluid or in the pseudogap phase the behavior of
EDC is more complicated as it contains the mixture of various dispersion relations. One can relate
the deviation from the $p^{2}$ behavior to the presence of other phases in the system.
The behavior of EDC at high $p$ corresponds to the occupied branch 
$A^{(-)}$ for momenta above the Fermi level. At unitarity this behavior can be attributed to the
relation $n(p)\propto 1/p^{4}$ at large $p$ \cite{tan,schneider}. 
It is worth noting that the recent radio frequency spectra analysis of a trapped unitary $^{6}$Li also
revealed the existence of three distinct phases in a trap below $T_{c}$, 
similarly like in the upper panel of Fig. \ref{edc_fig} \cite{pieri}.

\end{document}